\documentclass{pasj01}
\usepackage{lineno}
\Received{$\langle$reception date$\rangle$}
\Accepted{$\langle$acception date$\rangle$}
\Published{$\langle$publication date$\rangle$}

\begin{document}

\title{Suzaku Observations of Fe K-shell Lines in the Supernova Remnant W51C and Hard X-ray Sources in the Proximity}
\author{Aika  \textsc{Shimaguchi}\altaffilmark{1}, Kumiko K. \textsc{Nobukawa}\altaffilmark{2,1}\altaffilmark{${\ast}$},   Shigeo \textsc{Yamauchi}\altaffilmark{1}, Masayoshi  \textsc{Nobukawa}\altaffilmark{3}, and Yutaka \textsc{Fujita}\altaffilmark{4}}
\altaffiltext{1}{Faculty of Science, Nara Women's University, Kitauoyanishi-machi, Nara, Nara 630-8506, Japan}
\altaffiltext{2}{Faculty of Science and Engineering, Kindai University, 3-4-1 Kowakae, Higashi-Osaka, 577-8502, Japan}
\altaffiltext{3}{Department of Teacher Training and School Education, Nara University of Education, Takabatake-cho, Nara, 630-8528, Japan}
\altaffiltext{4}{Department of Physics, Graduate School of Science, Tokyo Metropolitan University, 1-1 Minami-Osawa, Hachioji-shi, Tokyo 192-0397, Japan}
\email{kumiko@phys.kindai.ac.jp}


\maketitle

\begin{abstract}
In this paper, we investigated the Fe K-shell lines in the supernova remnant W51C and hard X-ray sources in the proximity. 
We measured the intensities of Fe\emissiontype{I} K$\alpha$ and Fe\emissiontype{XXV} He$\alpha$ lines at 6.40~keV and 6.68~keV, respectively,  and found that the intensity of the 6.68~keV line is consistent with the background level expected from previous studies, while that of the 6.40~keV line is higher at the significance level of $2.0\sigma$. Given the presence of gamma-ray emission and high ionization rate point spatially coincident with the remnant, we conclude that the enhanced 6.40~keV line most likely originates from the interaction between low-energy cosmic rays and molecular clouds.
Also, we discovered an enhanced 6.68~keV line emission from the compact H\emissiontype{II} region G49.0$-$0.3  at the significance level of $3.4\sigma$.  Spectral analysis revealed that the temperature and abundance of the thermal plasma with the 6.68~keV line is $kT=3.0^{+0.8}_{-0.7}$~keV and $Z= 0.5\pm 0.2$~solar, respectively. These values are explained by the thermal plasma generated by the stellar winds of O stars. 
\end{abstract}

\section{Introduction}
W51C is a middle-aged supernova remnant (SNR) located at a distance of $D\simeq6$~kpc \citep{Koo05} toward the direction of $(l, b) = (\timeform{49D.2}, \timeform{-0.D7})$, which corresponds to the tangential point of the Sagittarius arm. The SNR is superposed on the massive star-forming region W51B on its western side,  where there are found shocked atomic H\,\emissiontype{I} gas \citep{Koo91, Koo97b} (see figure~1) and molecules such as CO and HCO$^+$ \citep{Koo97c}, and also OH (1720 MHz) masers \citep{Green97}. 

Gamma-ray emission was also found from the W51C region by the Fermi/LAT \citep{Abdo09}, H.E.S.S. \citep{Fiasson09}, and MAGIC \citep{Aleksic12}. The luminosity of the GeV gamma-ray emission is estimated to be  $\simeq1\times10^{36}(D/6$~kpc$)^2$~erg~s$^{-1}$ (where $D$ is the distance to W51C), which makes the SNR one of the most luminous Galactic sources in gamma rays. Since the gamma-ray spectrum is well explained by $\pi^0$ decays \citep{Abdo09, Aleksic12},  gamma rays could be plausibly produced by interactions between high-energy protons and clouds. The high ionization rate of $\zeta\sim10^{-15}$~s$^{-1}$ was estimated based on a measurement of the DCO$^+/$HCO$^+$ abundance ratio \citep{Ceccarelli11}.  Those results suggest the existence of a large number of cosmic rays (CRs) accelerated at the shock.  

X-rays from W51C have been observed since the detection by the Einstein satellite \citep{Seward90}. ROSAT, ASCA, and Chandra revealed that its spectra were represented by a collisional ionization equilibrium (CIE) plasma with $kT\sim0.3$~keV \citep{Koo95, Koo02, Koo05}, whereas the spectra obtained by Suzaku and XMM-Newton were explained by a non-equilibrium ionization (NEI) model with $kT\sim0.6$--0.7~keV \citep{Hanabata13, Sasaki14}. The low-$kT$ values indicate that an X-ray spectrum of W51C is dominated by the soft band. A drop of the surface brightness in the soft X-ray band is found toward the SNR's western side;  it would be due to the absorption and thus the X-rays are emitted behind molecular clouds \citep{Koo95}.    

As for the hard X-ray band, ASCA discovered point-like sources (HX1, HX2, HX3east, and HX3west) \citep{Koo02}. Chandra revealed that HX2 is a pulsar wind nebula (PWN), and the source is named CXO\,J192318.5$+$1403035 \citep{Koo05}. 
HX1 and HX3east coincide with compact H\,\emissiontype{II} regions (G48.9$-$0.3 and G49.0$-$0.3 for HX1, and G49.2$-$0.3 for HX3 east), which belong to W51B. X-rays from HX1 and HX3east might come from young massive stars in these compact H\,\emissiontype{II} regions \citep{Koo02, Sasaki14}.  \citet{Sasaki14} found that HX3west has an X-ray spectrum similar to that of CXO\,J192318.5$+$140303,  and classified the source as a PWN candidate. \citet{Hanabata13} reported an extended hard X-ray emission from the W51B region. Its spectrum was represented by a power-law function with a photon index of $\sim2.2$ or a thermal plasma model with $kT\sim5$~keV. The authors claimed that X-rays come from gas heated at stellar shocks of OB-type stars or non-thermal emission by particles accelerated at the shocks. 

 A K-shell line from He-like Fe ions at $E=6.68$~keV indicates the existence of a high-temperature plasma with $ kT\gtrsim1$~keV.  The 6.68~keV line has been detected from some SNR plasmas (e.g., \cite{Yamaguchi14}). Also star-forming region can be  accompanied with the 6.68~keV line because shocked stellar winds reach to temperature of $kT=5$~keV$(v/2000~\rm{km~s}^{-1})^2$ behind the shock (e.g. \cite{Ezoe06}). In fact, some star-forming regions exhibit the 6.68~keV line \citep{Ezoe06, Sawada09}. 
If the hard X-rays from the star-forming region W51B or the H\,\emissiontype{II} regions therein are of thermal origin, they represent the 6.68~keV line.  Also, W51C, the SNR itself, can exhibit the 6.68~keV line if it has an undetected high-temperature ($kT\gtrsim1$~keV) plasma.

Fluorescence from neutral Fe atoms at $E=6.40$~keV is a probe of low-energy CRs (protons in the MeV band and electrons in the keV band). Actually, the 6.40~keV line has been observed from several SNRs \citep{Sato14, Sato16, Bamba18, Nobukawa18, Saji18, Nobukawa19}. The low-energy CRs do not diffuse much compared with the high-energy CRs, and therefore the accelerated low-energy particles interact with the interstellar gas very close to the acceleration site and emit the 6.40~keV line \citep{Makino19}. Since W51C interacts with molecular clouds and shows the gamma-ray emission and the high ionization rate,  there are potentially a large number of low-energy CRs interacting with clouds, which can result in an enhancement of the 6.40~keV line \citep{Fujita21}.

In this paper,  we focused on the Fe K-shell lines in W51C and hard X-ray sources in the proximity, and searched for undetected high-temperature plasma and the evidence for the low-energy CRs. The W51C region is on the Galactic plane, and therefore a major background is the Galactic ridge X-ray emission (GRXE),  which is also known to be accompanied by Fe K-shell lines \citep{Koyama18}. We evaluated the GRXE contribution and found a hint of an enhancement of the 6.40~keV line over the GRXE in the W51C region. We also report a locally enhanced 6.68~keV line in the compact H\,\emissiontype{II} region G49.0$-$0.3.

\section{Observations and Data Reduction}\label{sec:Obs}
We utilized the Suzaku archive data of two observations of W51C (OBSID: 504066010 and 504067010). Observation logs are shown in table~\ref{obserlog}.  
We used the X-ray Imaging Spectrometer (XIS; \cite{Koyama07}) data. The XIS consists of the four X-ray CCD cameras (XIS0, 1, 2, and 3), which are placed on the focal plane of the X-ray Telescope (XRT; \cite{Serlemitsos07}). XIS0, 2, and 3 employ Front-Illuminated (FI) CCDs, while XIS1 does a Back-Illuminated (BI) CCD. A field of view (FOV) of each CCD is $17'.8\times17'.8$. Since the entire region of XIS2 and the quarter of XIS0 had been out of function since 2006 November and 2009 June, respectively, data from these regions were not used.

We used the final Suzaku archive data which were re-processed with the latest software version and calibration information.   Data analysis was performed using the HEADAS software of version 6.28 and the calibration database (CALDB) released in February 2016. The redistribution matrix file (RMF) and the ancillary response file (ARF; \cite{Ishisaki07}) for spectral analysis were created by the \texttt{xisrmfgen} and \texttt{xissimarfgen} tools, respectively. The non-Xray Background (NXB) was estimated by \texttt{xisnxbgen} \citep{Tawa08}. Errors quoted in this paper and error bars given in the figures show the 1$\sigma$ statistical errors.

\begin{table*}[htb]
 \caption{Observation logs.}
 \centering
 \begin{tabular}{cccccc} \hline
  Region  &  Sequence No. & Galactic coordinate  & \multicolumn{2}{c}{Observation time (UT)} & Exposure  \\ 
  		&		& 	$( l, b )$		& Start 	& End 	& (ks)	     \\ \hline
  W51C west & 504066010 & (\timeform{49D.11}, \timeform{-0D.31})  & 2010-03-28 11:44:41 & 2010-03-29 13:19:12 & 44.1\\ 
  W51C east  & 504067010 & (\timeform{49D.11}, \timeform{-0D.54}) & 2010-03-30 07:23:48 & 2010-03-31 09:00:20 & 43.7\\
 \hline
  \end{tabular}
 \label{obserlog}
\end{table*}

\section{Analysis and Results}
\subsection{Intensities of Fe K-shell Lines of the Whole W51C Region}\label{sec:Fe2}
Figure 1(a) shows an X-ray image obtained by Suzaku in the 4--8~keV band (color) overlaid by the 0.5--2~keV band image (yellow contours). Here the images of XIS0, 1, and 3 were merged. The NXB was subtracted and the vignetting was corrected.  The outermost white line indicates the FOVs of the two observations. 
The hard X-ray objects previously reported, HX1, HX2 (CXO\,J192318.5$+$1403035), HX3east, and HX3west,  were seen in the image.  Soft X-ray emission would come from the SNR W51C. 
Figure 1(b) is the same as (a), but the yellow contours indicate the 1.4~GHz radio continuum \citep{Stil06, Beuther16}. The radio shell covers the whole FOVs. Since CR acceleration occurs in the shell, the 6.40~keV line should be searched in the whole FOVs. 

\begin{figure*}
 \begin{center}
  \includegraphics[width=17cm]{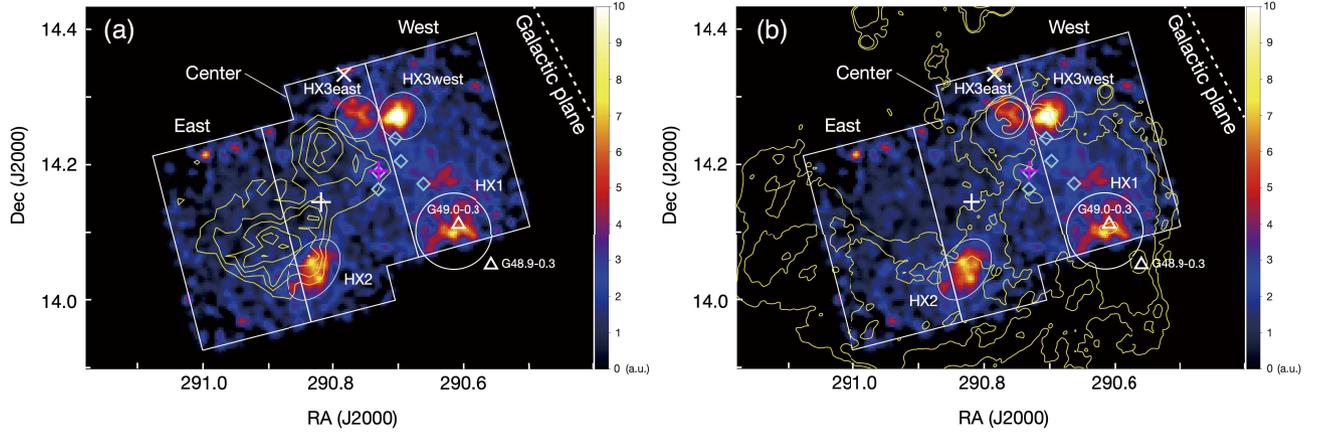}
 \end{center}
 \caption{(a) Hard X-ray image of the W51C region obtained by Suzaku in the 4--8~keV band (color). 
 The yellow contours indicate the 0.5--2~keV band image. The NXB was subtracted and the vignetting was corrected. 
 Circles and ellipses labeled HX1, HX2, HX3east, and HX3west represent hard X-ray sources. Spectra of  Center, West, and East were extracted from the regions excluding the circles and ellipses. White triangles indicate  H\emissiontype{II} regions found near HX1. Cyan diamonds indicate the positions of the shocked CO clouds \citep{Koo97c}.  "X" mark indicates the position where the high ionization rate was observed (position E in \cite{Ceccarelli11}) while "+" marks represent the centroids of gamma-ray emissions observed by MAGIC (magenta; \cite{Aleksic12}) and  Fermi (white; \cite{Abdo09}), respectively. Dashed line indicates a Galactic plane with $b=\timeform{-0D.046}$. (b) Same as (a), but the yellow contours indicate the 1.4~GHz radio continuum combined with the VGPS data \citep{Stil06, Beuther16} with a level from 0.045  to 2.0~Jy~beam$^{-1}$ with logarithmic step.}\label{fig:img}
\end{figure*}

We extracted a spectrum from the whole FOVs, using the FI and BI data. 
Here, the regions including the four hard X-ray objects (the circles and ellipses in figure 1) were excluded.
We merged the FI (XIS0 and 3) spectra and fitted the FI and BI spectra simultaneously. We fitted the spectrum of the high-energy band above 4.5~keV in order to avoid contamination of a local soft X-ray emission from W51C  (c.f. \cite{Hanabata13}; \cite{Sasaki14}; also see figure~\ref{fig:img}). We ignored the band above 10 keV for the FI spectrum and the band above 7.0 keV for the BI spectrum because the influence of the NXB is significant.    

The fitted model was an absorbed thermal bremsstrahlung plus four Gaussians in addition to the cosmic X-ray background (CXB). We used \texttt{phabs} in XSPEC as the absorption model. 
The thermal bremsstrahlung is mainly due to the GRXE and the extended hard X-ray emission overlapping W51B \citep{Hanabata13}, as well as a very small amount of thermal emission from W51C. The contribution of the thermal emission from W51C is only $\sim2$\% in the 4.5--10~keV band because of its low temperature ($\sim0.5$~keV; \cite{Hanabata13, Sasaki14}).
The four Gaussians represent the Fe K-shell lines, and the centroids were fixed to 6.40~keV and 7.05~keV (Fe\emissiontype{I} K$\alpha$ and K$\beta$), 6.68~keV (Fe\emissiontype{XXV} He$\alpha$), and  6.97~keV (Fe\emissiontype{XXVI} Ly$\alpha$). The normalization  of the Fe\emissiontype{I} K$\beta$ was fixed to 0.125 times that of  the Fe\emissiontype{I} K$\alpha$ \citep{Kaastra93, Smith01}. The parameters of the CXB were fixed to the values  in \citet{Kushino02}.  

\begin{figure}
 \begin{center}
  \includegraphics[width=8cm]{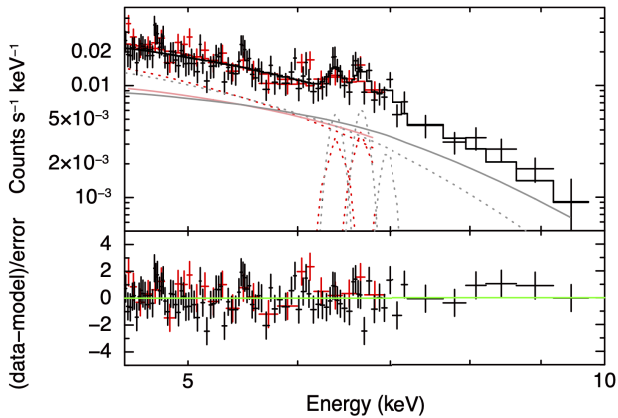}
 \end{center}
 \caption{Spectra of the W51C region in the 4.5--10 keV band. The FI (black) and BI (red) spectra were simultaneously fitted. Black and red sold lines are the best-fit model,  which consists of  the Fe\emissiontype{I} K$\alpha$ and K$\beta$,  Fe\emissiontype{XXV} He$\alpha$, and Fe\emissiontype{XXVI} Ly$\alpha$ lines plus  thermal bremsstrahlung (dotted lines) as well as the CXB (gray and light red solid lines). }\label{fig:spec}
\end{figure}

The best-fit parameters are summarized in table~\ref{tab:Feline1}. The intensities of the 6.40~keV and 6.68~keV lines were obtained to be $(1.4\pm0.4)\times 10^{-8}$ and $(1.7\pm0.4)\times 10^{-8}$  in units of photons~s$^{-1}$~cm$^{-2}$~arcmin$^{-2}$, respectively.  The spectrum with the best-fit model is presented in figure~\ref{fig:spec}.  
In the case where the centroid of the 6.40~keV line was let free, its upper limit was obtained to be 6.430~keV at the 90\% confidence level. Allowing the CXB flux to vary within the range of fluctuation in \citet{Kushino02} does not affect the best-fit line intensity.

Since there is no Suzaku data that can be used as background regions in the vicinity of W51C, we estimated the intensities of Fe K-shell lines of the background,  based on the GRXE model of previous studies. The distributions of the 6.40~keV and 6.68~keV lines accompanied by the GRXE were well studied by \citet{Uchiyama13} and \citet{Yamauchi16}. 
The scale lengths along the Galactic plane of the 6.40~keV and 6.68~keV lines are $57^{\circ}\pm50^{\circ}$ and $45^{\circ}\pm10^{\circ}$, respectively, according to \citet{Uchiyama13}.   As for the scale heights,  \citet{Uchiyama13} used the data mainly near the Galactic center, and the scale-height measurement was affected by the Galactic Bulge X-ray Emission (c.f. \cite{Koyama18}). The scale heights of the GRXE reported by \citet{Yamauchi16}  are more reliable because they measured a pure GRXE in the range of $|l| = 10^{\circ}$--$30^{\circ}$.   Therefore, we adopted the scale heights in \citet{Yamauchi16}, $\timeform{0D.50}\pm\timeform{0D.12}$ and $\timeform{1D.02}\pm\timeform{0D.12}$ for the 6.40~keV and 6.68~keV lines, respectively.  
Here, we considered the statistical errors of the scale length and scale height of each line obtained by \citet{Uchiyama13} and \citet{Yamauchi16}. \citet{Uchiyama13}  calculated the differences between the GRXE model and the data for the individual FOVs of Suzaku and obtained their standard deviation to be 24\%--30\%. This uncertainty was also taken into account.

The line intensities  at the central coordinate of the W51C regions ($l=\timeform{49D.11}, b=\timeform{-0D.43}$)  were estimated to be $(0.4\pm0.3)\times10^{-8}$ and $(2.0\pm0.8)\times10^{-8}$ in units of photons~s$^{-1}$~cm$^{-2}$~arcmin$^{-2}$ for the 6.40~keV and 6.68~keV lines, respectively. 
The observed intensity of the 6.68~keV line was consistent with the model, but that of the 6.40~keV line is higher at the confidence level of 2.0$\sigma$. 
It indicates a hint of an enhancement of the 6.40~keV line emission in the W51C region.

\subsection{Intensities of Fe K-shell Lines of Hard X-ray Sources}
For spectral analysis of the hard X-ray sources (HX1, HX2, HX3east, and HX3west), we also extracted spectra from the circle and ellipse regions in figure~1, respectively, and fitted the individual spectra with an absorbed thermal bremsstrahlung and four Gaussians as well as the CXB.
The best-fit intensities of the 6.40~keV and 6.68~keV lines are shown in table~\ref{tab:Feline1}. The line intensities are consistent with those of the W51C region within the error ranges. The only exception is the 6.68~keV line of HX1, whose intensity is higher than that of the W51C region at the significance level of $3.4\sigma$.  According to \citet{Koo02}, HX1 is coincident with the compact H\emissiontype{II} regions G48.9$-$0.3 and G49.0$-$0.3. In our hard X-ray image (figure~\ref{fig:img}),  the emission peak of the HX1 region is very close to G49.0$-$0.3, and therefore the enhancement of the 6.68~keV line would come from G49.0$-$0.3.

\begin{table*}
  \tbl{The best-fit parameters of the whole region (except the hard X-ray sources), each hard X-ray source, West, Center, and East.}{%
  \centering
  \begin{tabular}{lcccccccc}
  \hline
Parameters 	&  \multicolumn{8}{c}{Regions} \\ \hline
  			& Whole region	&	HX1			& HX2			& HX3east 		& HX3west		& West			&	Center		& East	\\ \hline
$N_{\rm H} (10^{22}$~cm$^{-2})$	&	   \multicolumn{8}{c}{2.0 (fixed)$^{\ast}$} \\
$kT$ (keV)	& $3.6^{+1.1}_{-0.8}$		& $4.4^{+2.8}_{-1.4}$ 	& $5.3^{+2.5}_{-1.5}$	& $4.6^{+5.6}_{-1.9}$	& $>13.0$	& $4.0^{+1.6}_{-1.0}$	& $6.0^{+16.1}_{-2.9}$	& $3.4^{+4.7}_{-1.4}$	\\
Flux$_{\rm 4.5-10~keV}^{\dag}$ 	& $14.8\pm0.6$	&	$2.8\pm0.2$	&	$2.9\pm0.2$	&	$1.4\pm0.1$	&	$2.8\pm0.2$		&	$6.2\pm0.3$	 	&  $	5.4\pm0.5$	& $3.4\pm0.3$	 	\\
Intensity of the 6.40~keV line$^{\ddag}$	& $1.4\pm0.4$	&  $3.1\pm1.9$ 	& $4.3\pm3.0$ 	& 	$< 6.9$	& $2.8\pm2.5$	& $1.0\pm0.7$ 	& $1.6\pm0.6$	& $1.2\pm0.6$ \\
Intensity of the 6.68~keV line$^{\ddag}$	& $1.7\pm0.4$	&  $9.9\pm2.4$ 	& $<1.6$			& 	$<6.7$	& $<3.7$			& $1.9\pm0.5$ 	& $1.8\pm0.6$ 	& $<1.1$   \\
\hline
$\chi^2/$d.o.f. & $118.8/115$	& $43.7/57$	&  $53.2/57$	& $37.9/46$	& $40.4/57$	& $91.0/92$	& $85.1/82$	& $70.5/90$ \\
\hline
\multicolumn{9}{l}{$^{\ast}$ Fixed to the typical value in W51C \citep{Koo05, Hanabata13, Sasaki14}.} \\
\multicolumn{9}{l}{$^{\dag}$ Observed flux in the unit of $10^{-13}$~erg~s$^{-1}$~cm$^{-2}$ in the 4.5--10~keV band. The GRXE is included, but the CXB is not.} \\
\multicolumn{9}{l}{$^{\ddag}$ Unabsorbed intensity. Units are $10^{-8}$~photons~s$^{-1}$~cm$^{-2}$~arcmin$^{-2}$. } \\
   \end{tabular}}\label{tab:Feline1}
 \end{table*}

\subsection{Spatial Distribution of the Fe K-shell Lines}
To investigate the spatial distribution of the 6.40~keV line, we divided the W51C region into three regions:  West, Center, and East as shown in figure~\ref{fig:img}. A spectrum of the high-energy band  ($\geq4.5$~keV) was extracted from each region excluding the circle and ellipse regions in figure 1.  We fitted the spectra with an absorbed thermal bremsstrahlung plus four  Gaussians and the CXB. The obtained line intensities are summarized in table~\ref{tab:Feline1}.  
The intensities of the 6.40~keV line of the three regions are consistent within  $1\sigma$ errors, 
whereas that of the 6.68~keV line is weakest in East, where is the farthest region from the Galactic plane among the three regions. The GRXE weakens with a distance from the Galactic plane \citep{Uchiyama13, Yamauchi16}. The weak intensity of the 6.68~keV line in East is consistent with that expected from the GRXE distribution.

\subsection{Spectral analysis of  G49.0$-$0.3}
\begin{figure}
 \begin{center}
  \includegraphics[width=9cm]{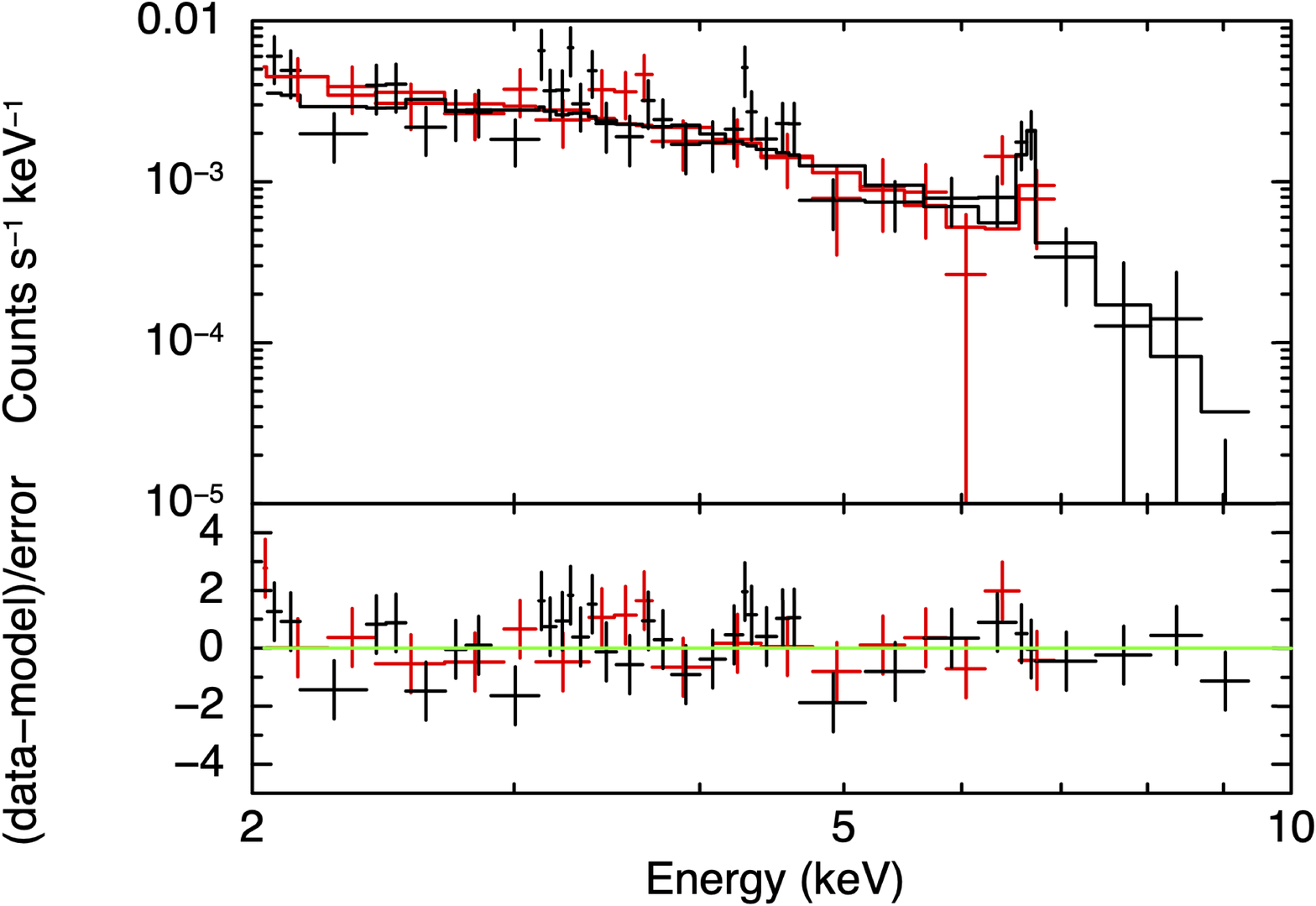}
 \end{center}
 \caption{Background-subtracted spectrum of G49.0$-$0.3.  The FI (black) and BI (red) spectra were simultaneously fitted. Black  and red solid line represents the best-fit model.}\label{fig:G49}
\end{figure}

The enhanced 6.68~keV line emission in  G49.0$-$0.3 (HX1)  implies the presence of high-temperature plasma. 
We fitted the G49.0$-$0.3 spectrum in the 2--10~keV band with an absorbed CIE plasma model ({\tt apec} in XSPEC). Here, the spectrum of West was used as the background spectrum. The model explained well the G49.0$-$0.3 spectrum ($\chi^2/$d.o.f. = 58/52 = 1.12). The temperature and the abundance of the CIE plasma were obtained to be $kT = 3.0^{+0.8}_{-0.7}$~keV and $0.5 \pm 0.2$~solar, respectively. The flux of the 2--10 keV band was $3.7 \times 10^{-13}$~erg~s$^{-1}$~cm$^{-2}$. The spectrum is shown in figure~\ref{fig:G49} and the best-fit parameters are shown in  table~\ref{G49.0}.

\begin{table*}
  \tbl{Best-fit parameters of the G49.0$-$0.3 spectrum.}{%
  \begin{tabular}{lc}
  \hline
Parameter	 					&  Value    \\  
  \hline
$N_{\rm H} (10^{22}$~cm$^{-2}$) 	& $2.5^{+1.2}_{-0.9}$ 	\\
$kT$ (keV)	 					& $3.0^{+0.8}_{-0.7}$    \\
Abundance (solar)					 & $0.5\pm0.2$ \\
Flux$_{\rm 2-10~keV}^{\ast}$  						& $3.7\pm0.2$ \\
  \hline
$\chi^2/$d.o.f.  						&  $58.0/52$ \\
  \hline
\multicolumn{2}{l}{$^{\ast}$  Observed flux in the unit of $10^{-13}$~erg~s$^{-1}$~cm$^{-2}$ in the 2--10~keV band. } \\
   \end{tabular}}\label{G49.0}
 \end{table*}

\section{Discussion}
 
\subsection{Origin of the Enhancement of the 6.68~keV Line  in G49.0$-$0.3}
We found the enhancement of the 6.68~keV line  in the compact H\emissiontype{II} region G49.0$-$0.3.  In H\emissiontype{II} regions, stellar winds experience a shock transition, and the shocked hot winds fill the shell region to emit thermal X-rays.  The shock temperature is expressed as follows:
\begin{equation}
kT  = \frac{3}{16}\mu m_{\rm H}v_w^2 = 5\, ( v_w /2000~{\rm km~s}^{-1})^2~{\rm keV},
\end{equation}
 where $m_{\rm H}$ is the mass of a hydrogen atom, $\mu = 0.62$ is the mean molecular weight, and $v_w$ is the stellar wind velocity (e.g. \cite{Ezoe06}). Since the observed temperature of the G49.0$-$0.3 region is $kT = 3.0^{+0.8}_{-0.7}$~keV,  the wind velocity is estimated to be $v_w\sim1500$~km~s$^{-1}$.  The most massive star in G49.0$-$0.3 is an O9 star \citep{Kim07}, and a typical wind velocity of O stars is 1000--3000~km~s$^{-1}$ \citep{Prinja90}. The observed abundance was $0.5\pm 0.3$~solar, which is consistent with the values in the other H\emissiontype{II} regions (e.g., \cite{Hyodo08}). Therefore, the enhancement of the 6.68~keV line would be due to the high-temperature plasma associated with the H\emissiontype{II} region.

\citet{Hanabata13} reported the extended hard X-ray emission overlapping W51B  (called ``Reg.3")  although the G49.0$-$0.3 region was excluded in their analysis. They found that the spectrum of the hard X-ray emission was represented by either a high-temperature thermal plasma model or a  non-thermal emission,  but they did not distinguish which model is better. ``Reg.3" coincides with a part of West and  Center regions. Our analysis found no enhancement of the 6.68 keV line compared to the GRXE in both regions, which supports the non-thermal scenario.

\subsection{Origin of the Enhancement of the 6.40~keV Line Emission}
We found the enhancement of the 6.40~keV line at the significance level of $2.0\sigma$ although it is not a robust result. Here we discuss possible origins of the 6.4 keV line enhancement, assuming that it is true.

One possible origin of the enhancement is the ionizing thermal plasma of the SNR. However, we conclude that the 6.40~keV line is not likely of the plasma origin. First, because the thermal plasma of W51C is in the ionizing state and its temperature is $kT\sim0.6$--$0.7$~keV \citep{Hanabata13, Sasaki14}, at which a detectable Fe K-shell line cannot be emitted.  Second because, as mentioned in section~\ref{sec:Fe2}, the upper limit of the centroid of the observed 6.40~keV line is 6.430~keV, which corresponds to the ionization state of Ne-like Fe and the ionization time-scale of $\sim200$~yr  according to the NEI  code in XSPEC. This is two orders of magnitudes younger than the age of W51C \citep{Koo95}. Here we assume the electron number density of plasma of 1~cm$^{-3}$.    

The 6.40~keV line can be generated by X-ray photoionization. The Fe K-shell ionization energy is 7.1~keV, and therefore soft X-ray sources such as W51C contribute little to the 6.40 keV emission. Candidates are the four hard X-ray sources, HX1 (G49.0$-$0.3), HX2, HX3west, and HX3east. The flux of the 6.4 keV line is calculated as
\begin{eqnarray}
&F_{\rm 6.4~keV}   =  \nonumber  \\
&\epsilon \left(\frac{\Omega}{4\pi}\right) \int^{\infty}_{\rm 7.1~keV} [1-{\rm exp} (-N_{\rm H} Z_{\rm Fe} \sigma_{\rm Fe}(E))] A(E) dE
\end{eqnarray}
where $\epsilon$  is the Fe fluorescence yield, $\Omega$ is a solid angle with which the molecular clouds are seen from the source, $N_{\rm H}$ is the hydrogen column density of the molecular clouds, $Z_{\rm Fe}$ is the Fe abundance,  $\sigma_{\rm Fe}(E)$ is the cross-section of the photoionization for Fe atoms, and $A(E) = K(E/1~{\rm keV})^{-\Gamma}$ is the X-ray spectrum of the irradiating X-ray source with the photon index $\Gamma$ and normalization $K$. Here, we used  the upper limits of the hydrogen column densities $N_{\rm H}$  of the shocked CO and H \emissiontype{I} of $N_{\rm H} \geq 4.0\times10^{16}$~cm$^{-2}$ and $N_{\rm H} \geq 1.8\times10^{21}$~cm$^{-2}$, respectively \citep{Koo97b, Koo97c}.  Also, we adopted $\epsilon = 0.34$ \citep{Bambynek72}, $Z_{\rm Fe}=3\times10^{-5}$  \citep{Lodders03}, and $\sigma_{\rm Fe}(E)=6\times10^{-18} (E/1~{\rm keV})^{-2.6}$~cm$^{2}$ \citep{Henke82}.   The photon index $\Gamma$ of the four hard X-ray sources ranges 1.8--3.2 (\cite{Sasaki14}, this work) and  the flux of the enhanced 6.40~keV line we obtained in this work is  $F_{\rm 6.4~keV} = 5.0\times10^{-6}$~photons~s$^{-1}$~cm$^{-2}$. 
Substituting these values into equation (2), we obtained the normalization $K$ of $A(E)$ to be $(0.1$--$3) (4\pi/\Omega)(1.8\times 10^{21}~{\rm cm}^{-2}/N_{\rm H})$~photons~s$^{-1}$~cm$^{-2}$~keV$^{-1}$. Then, integrating $A(E)$, we estimated the required X-ray flux of the irradiating source to be $F_X = (0.07$--$1)\times10^{-8}(4\pi/\Omega)(1.8\times 10^{21}~{\rm cm}^{-2}/N_{\rm H})$~erg~s$^{-1}$~cm$^{-2}$ in the 0.3--10 keV band,   which  is at least two orders of magnitude higher than the typical flux of the hard X-ray sources; $\sim 10^{-13}$~erg~s$^{-1}$~cm$^{-2}$ for HX1 (this work) and $\sim 10^{-12}$~erg~s$^{-1}$~cm$^{-2}$ for the other sources \citep{Sasaki14} regardless of the solid angle. It is unlikely that the 6.40~keV line is of the photoionizaiton origin.   

The most probable origin of the enhancement of the 6.40~keV line would be the interaction between the low-energy CRs and molecular clouds. 
In fact, W51C is thought to be a CR accelerator;  it is one of the brightest gamma-ray SNRs in the Galaxy \citep{Ajello20},  and the high ionization rate was measured  \citep{Ceccarelli11}. 
Previous observations have shown that there are shocked CO clouds at the boundary between Center and West, and the centroids of the gamma-ray emissions and the position of the high ionization rate are consistent with Center (figure~\ref{fig:img}). 

\citet{Makino19} demonstrated that both the 6.40~keV line and gamma-ray emissions produced by the proton-proton interaction can be explained by a CR escaping model for SNRs interacting with molecular clouds.   Assuming the 6.40~keV line in the W51C region is of the proton origin, we applied this model to W51C and found that the 6.40~keV line intensity calculated from the information of gamma-ray spectrum \citep{Aleksic12} and molecular clouds \citep{Carpenter98} is $0.3\times10^{-8}$~photons~s$^{-1}$~cm$^{-2}$~arcmin$^{-2}$, which is smaller than the observed value.  However, this may reflect that low-energy CRs are being further accelerated behind the shock of the SNR through the interaction between the shock and the clouds \citep{Inoue12}.

\begin{figure}
 \begin{center}
  \includegraphics[width=8cm]{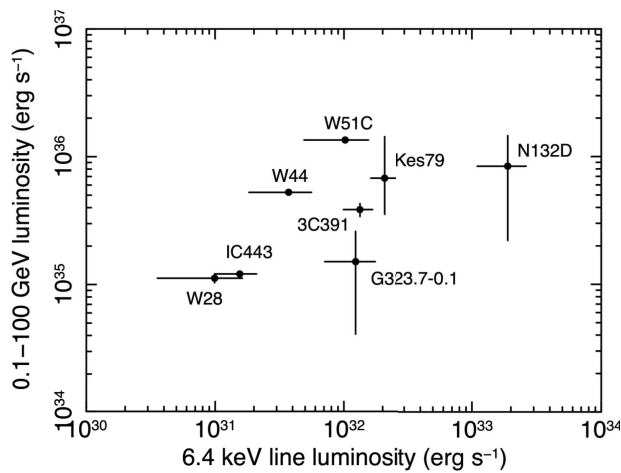}
 \end{center}
 \caption{Comparison between the 6.40~keV line luminosity and 0.1--100 GeV gamma-ray luminosity for 3C391 \citep{Sato14},  Kes\,79 \citep{Sato16}, N132D \citep{Bamba18}, W28, W44 \citep{Nobukawa18},  G323.7$-$0.1 \citep{Saji18}, IC\,443 \citep{Nobukawa19}, and W51C (this work). The 0.1--100~GeV gamma-ray luminosities are taken from the Fermi LAT 10-Year Point Source Catalog \citep{Ajello20}. }\label{fig:FeG}
\end{figure}

If the 6.40~keV line emission originates from low-energy protons interacting with molecular clouds, we can expect a positive correlation between the luminosities of the 6.40 keV line and gamma-rays. Besides W51C, there are several SNRs where both the 6.40~keV line and gamma-ray emission have been observed: 3C391,  Kes\,79, N132D, W28, W44,  G323.7$-$0.1, and  IC\,443 (\cite{Sato14, Sato16, Bamba18, Nobukawa18, Saji18, Nobukawa19}; \cite{Ajello20}). It would be valuable to investigate the relation between the 6.40~keV line and 0.1--100~GeV gamma-ray luminosities. As shown in figure~\ref{fig:FeG}, we found a hint of a  positive correlation between the two observables (the correlation coefficient is $r\sim0.3$) although the sample size is limited.

\section{Conclusion}
We investigated the Fe K-shell lines in W51C  with Suzaku. Since W51C is located near the Galactic plane, a major background is the GRXE, which is also accompanied by the Fe K-shell lines. We estimated the Fe K-shell line intensities accompanied by  the GRXE in the W51C region based on the previous studies and found that the observed 6.40~keV line intensity is higher than that of the GRXE with the significance level of 2.0$\sigma$. 
The most likely origin of the enhancement of the 6.40~keV line is the interaction between low-energy CRs and molecular clouds.  The existence or non-existence of the 6.40~keV line enhancement will be conclusively revealed by future high-resolution spectroscopic observations.

The observed 6.68~keV line intensity in the W51C region is consistent with that of the GRXE, but the local enhancement of the 6.68~keV line in the compact H\emissiontype{II} region G49.0$-$0.3 was found for the first time at the significance level of $3.4\sigma$.   Spectral analysis revealed that the temperature of the thermal plasma with the 6.68~keV line is $kT=3.0^{+0.8}_{-0.7}$~keV and abundance is $0.5\pm 0.2$~solar. These values would be explained by the thermal plasma produced by the stellar winds of O stars. 

\section*{Acknowledgements}
This work was supported by MEXT KAKENHI No.JP20K14491, JP20KK0071 (KN), and JP21K03615 (MN). KKN was also supported by Yamada Science Foundation.

\end{document}